\documentclass[twocolumn,10pt]{article} 

\usepackage[square,numbers,sort&compress,comma]{natbib}

\usepackage{amsmath}
\usepackage{amssymb}
\usepackage{caption}
\usepackage{graphicx}
\usepackage{latexsym}
\usepackage{times}
\usepackage[pagewise]{lineno}
\usepackage{tabularx}
\usepackage{subfig}

\topmargin - 12pt 
\renewenvironment{abstract}%
              {
               \small
               {\bfseries \abstractname}
               \par
               \vspace{10pt}
              }

\renewcommand\abstractname{Abstract}

\newcommand{\nomenclature}
              [1]
              {
               \bgroup
               \flushleft
               \small\bf
               #1
               \par
               \egroup
              }

\renewcommand{\section}
              [1]
              {
               \bgroup
               \flushleft
               \small\bf
               \refstepcounter{section}
               \arabic{section}. #1
               \par
               \egroup
              }

\renewcommand{\subsection}
              [1]
              {
               \bgroup
               \flushleft
               \small\em
               \refstepcounter{subsection}
               \arabic{section}.
               \arabic{subsection}. #1
               \par
               \egroup
              }

\renewcommand{\subsubsection}
              [1]
              {
               \bgroup
               \flushleft
               \small\em
               \refstepcounter{subsubsection}
               \arabic{section}.
               \arabic{subsection}.
               \arabic{subsubsection}. #1
               \par
               \egroup
              }

  \newcommand{\acknowledgement}
              [1]
              {
               \bgroup
               \flushleft
               \small\bf
               #1
               \par
               \egroup
              }

  \newcommand{\sectionbib}
              [1]
              {
               \bgroup
               \flushleft
               \small\bf
               #1
               \par
               \egroup
              }

\begin{document}

\title{\LARGE Variational Auto-Encoder Based Deep Learning Technique For Filling Gaps in Reacting PIV Data}

\author{{\large Shashank Yellapantula$^{a,*}$}\\[10pt]
        {\footnotesize \em $^a$National Renewable Energy Laboratory, Golden, CO, USA}\\[-5pt]
        }

\date{}


\small
\baselineskip 10pt


\twocolumn[\begin{@twocolumnfalse}
\vspace{50pt}
\maketitle
\vspace{40pt}
\rule{\textwidth}{0.5pt}
\begin{abstract} 
In this study, a deep learning based conditional density estimation technique known as conditional variational auto-encoder (CVAE) is used to fill gaps typically observed in particle image velocimetry (PIV) measurements in combustion systems. The proposed CVAE technique is trained using time resolved gappy PIV fields, typically observed in industrially relevant combustors. Stereo-PIV (SPIV) data from a swirl combustor with very a high vector yield is used to showcase the accuracy of the proposed CVAE technique. Various error metrics evaluated on the reconstructed velocity field in the gaps are presented from data sets corresponding to three sets of combustor operating conditions. In addition to accurate data reproduction, the proposed CVAE technique offers data compression by reducing the latent space dimension, enabling the efficient processing of large-scale PIV data.
%
\end{abstract}
\vspace{10pt}
\parbox{1.0\textwidth}{\footnotesize {\em Keywords:} VAE; PIV; Deep Learning; CNN; Combustor}
\rule{\textwidth}{0.5pt}
\vspace{10pt}

\end{@twocolumnfalse}] 

\clearpage

\twocolumn[\begin{@twocolumnfalse}

\centerline{\bf Information for Colloquium Chairs and Cochairs, Editors, and Reviewers}

\vspace{20pt}



\vspace{20pt}

{\bf 1) Novelty and Significance Statement}
\vspace{10pt}


\vspace{10pt}
 The following are the novel contributions made by this study to advance techniques for accurately filling large gaps in reacting PIV data:  
\begin{enumerate}
\item First of a kind application of a generative ML technique to fill gaps in reacting PIV data. Technique solely relies on time varying vector field in the valid regions. 
\item Usage of partial convolution operator to map spatially fine flow features, solely in the valid regions, to network parameters.
\item Novel use of conditioning variables to uniquely identify and reproduce distinct flow features in each of the snapshots. 
\item Demonstrate the superior accuracy of CVAE in reproducing vector field in the gaps even in extreme cases where 75\% of the vectors were eliminated from the training dataset. 
\end{enumerate}

 This paper presents significant research that addresses a long-standing challenge in processing gaps in large reacting PIV datasets commonly found in industrially relevant combustion systems.

\vspace{20pt} 

{\bf 2) Author Contributions}
\vspace{10pt}

The first and the only author designed and performed the research along with writing and editing of the paper.

\vspace{10pt}

{\bf 3) Authors' Preference and Justification for Mode of Presentation at the Symposium}
\vspace{10pt} 


\vspace{10pt}

The author prefers {\bf OPP} presentation at the Symposium, for the following reasons:

\begin{itemize}

  \item{This study will benefit both the experimentalists and the modelers due to the accurate reconstruction and data compression capability.}

  \item{Presentation of a novel algorithm not applied previously in the experimental diagnostics community.}

  \item{This study presents a significant improvement over the most popular algorithms for reacting PIV data processing.}

\end{itemize}

\end{@twocolumnfalse}] 


\clearpage


\section{Introduction\label{sec:introduction}} \addvspace{10pt}

Particle image velocimetry (PIV) measurements in complex combustor flows often suffer from low signal to noise ratio. This can be attributed to several factors including flame luminosity, wide density variations and complex flow structures leading to non-uniform particle distributions that lead to non-uniform particle distributions. These factors can cause gaps in the velocity vector field. This issue is further exacerbated at high pressures representative of complex industrial combustion systems. Several techniques, including gappy proper orthogonal decomposition (GPOD)~\cite{saini2016development} have been proposed in literature to reconstruct the velocity field in these gaps. In this above referenced study, authors proposed a median filter (MF) based GPOD technique and compared the superior performance of MF GPOD method against some of the existing techniques such as the method by Gunes~\cite{gunes2006gappy}, Raben~\cite{raben2012adaptive} and the original GPOD algorithm~\cite{everson1995karhunen}. 
These techniques have been shown to perform very well in certain conditions. However, they seem to struggle for cases where large chunks of data were missing. Recently, studies focusing on improving the GPOD techniques using spatial and temporal correlations have been proposed~\cite{nekkanti2023gappy}. Additionally, some machine learning based techniques have also been proposed~\cite{morimoto2021experimental} where the authors used supervised machine learning techniques trained on artificial particle images (APIs) and direct numerical simulation (DNS) data to develop new data reconstruction method for PIV. Another very recent study~\cite{zhao2023unified} proposed a technique that unifies deep neural networks and GPOD for fluid and thermal reconstruction. 

In the current study a deep learning-based generative technique called the variational autoencoder (VAE) is proposed for PIV gap filling. This technique leverages Bayesian inference to effectively handle large training data with large gaps in vector field and achieve superior reconstruction accuracy. This flavor of VAE with conditioning data, also known as Conditional Variational Auto-Encoder (CVAE), is trained on the temporally evolving gappy vector field, enabling it to infer the reconstructed field in the gaps. This approach outperforms the above mentioned techniques by demonstrating higher reconstruction accuracy even in extreme cases where 75\% of vectors are eliminated from the training snapshots.
 
\section{Methodology\label{sec:method}} \addvspace{10pt}
Gappy POD~\cite{saini2016development}, as with all other forms of POD, is a dimension reduction technique with both an encoding and a decoding component. In the POD encoder stage, the high-dimensional data is projected onto the POD modes. This projection results in coefficients that provide the contribution of each mode. These coefficients form the latent space which is the low-dimensional representation of the original high-dimensional data. During the decoding phase, linear combination of these POD modes are employed to reconstruct the original high-dimensional data. By varying the number of coefficients and associated POD modes, the original data can be approximated with varying levels of accuracy. Variational auto-encoder is a similar although a non-linear approach to dimension reduction with some specific properties imposed on the latent space. 
\subsection{Conditional Variational Auto-encoder (CVAE)}
\label{sec:cvae}
In the current study, a generative deep Learning Model known as Variational auto-encoder (VAE)~\cite{kingma2013auto} is used to learn and predict the background distribution of the velocity field and provide an approximation to the velocity field in the gaps. A comprehensive and a detailed description of underlying variational and the deep learning techniques are beyond the scope of this article and the reader is referred to~\cite{kingma2013auto} and~\cite{sohn2015learning} for a detailed discussion. In the following a brief description of how these methods are implemented in the context of development of a technique to fill gaps in reactive PIV data will be presented. 

Traditionally, autoencoders (AE) have been used for dimension reduction and therefore for data compression purposes. In this regard, AE is analogous to POD. One critical difference between AE and POD is that AE is an inherently non-linear technique with both encoder and decoder typically represented by deep neural networks, fully connected layers and non-linear activation functions. The non-linear capability within AE allows this technique to learn representations of the input data in lower dimensions with minimal reconstruction loss. However, the latent spaces with AE are low dimensional vector spaces lacking any regularization, not necessarily continuous and without any ability to interpolate within the latent space. Lack of interpolation property within the latent space leads to AE reconstructing incorrect data if the latent space regions away from regions corresponding to input data are fed to the decoder network. For the current use case of training datasets with gaps, AE latent space will likely suffer from information loss regarding data in the gaps. This constraint with AE is removed in VAEs where the high-dimensional input data is encoded to a distribution over the latent space. Thereafter, the latent space distribution is sampled and decoded to generate an approximation to the original high-dimensional data. A regularization term, imposing constraints of completeness and continuity over the latent space, is added to the mean square reconstruction loss between the original and the decoded data. In practice, this regularization is implemented by enforcing latent space distributions to be close to a high-dimensional gaussian distribution with unit norm. In order to compute the parameters of the encoder and the decoder network this loss function is backpropagated using stochastic gradient descent. To allow for gradient descent through a distribution with a random variable, the reparameterization trick is applied~\cite{kingma2013auto}. The loss function can be summarized as the following 
\begin{equation}
    \mathcal{L} = -KL\left[\mathcal{N}(\mathbf{\mu}_x, \mathbf{\sigma}_x)|| \mathcal{N}(0,I)\right] + MSE(\mathbf{x}, \mathbf{x}_d)
    \label{eq:lossfunc}
\end{equation}
where $\textit{KL}$ is the Kullback-Leibler divergence, MSE is mean square error, $\mathbf{x}$ is the original data and  $\mathbf{x}_d$ is the decoded data. 

One novel step introduced in this study is to condition every snapshot in the input dataset (temporally resolved vector field with gaps at every time instant) with a few specific variables computed using the gappy field at that time instant. Conditioning variables are generated to control the output data at every instant and generate the most plausible vector field in the gaps taking into account specific characteristics in the training snapshot at that corresponding instant. This flavor of VAE is commonly referred to as conditional variational auto-encoder (CVAE)~\cite{sohn2015learning}. During the training process these conditioning variables are input to both the encoder and decoder. During the prediction stage these conditioning variables along with a sample from the latent space distribution are fed to the decoder to generate approximate samples of full vector field at every instant.

The next section delves into implementation details of each of the component of the CVAE employed in this study.
\begin{figure*}[ht]
    \centering
    \includegraphics[width=0.9\textwidth]{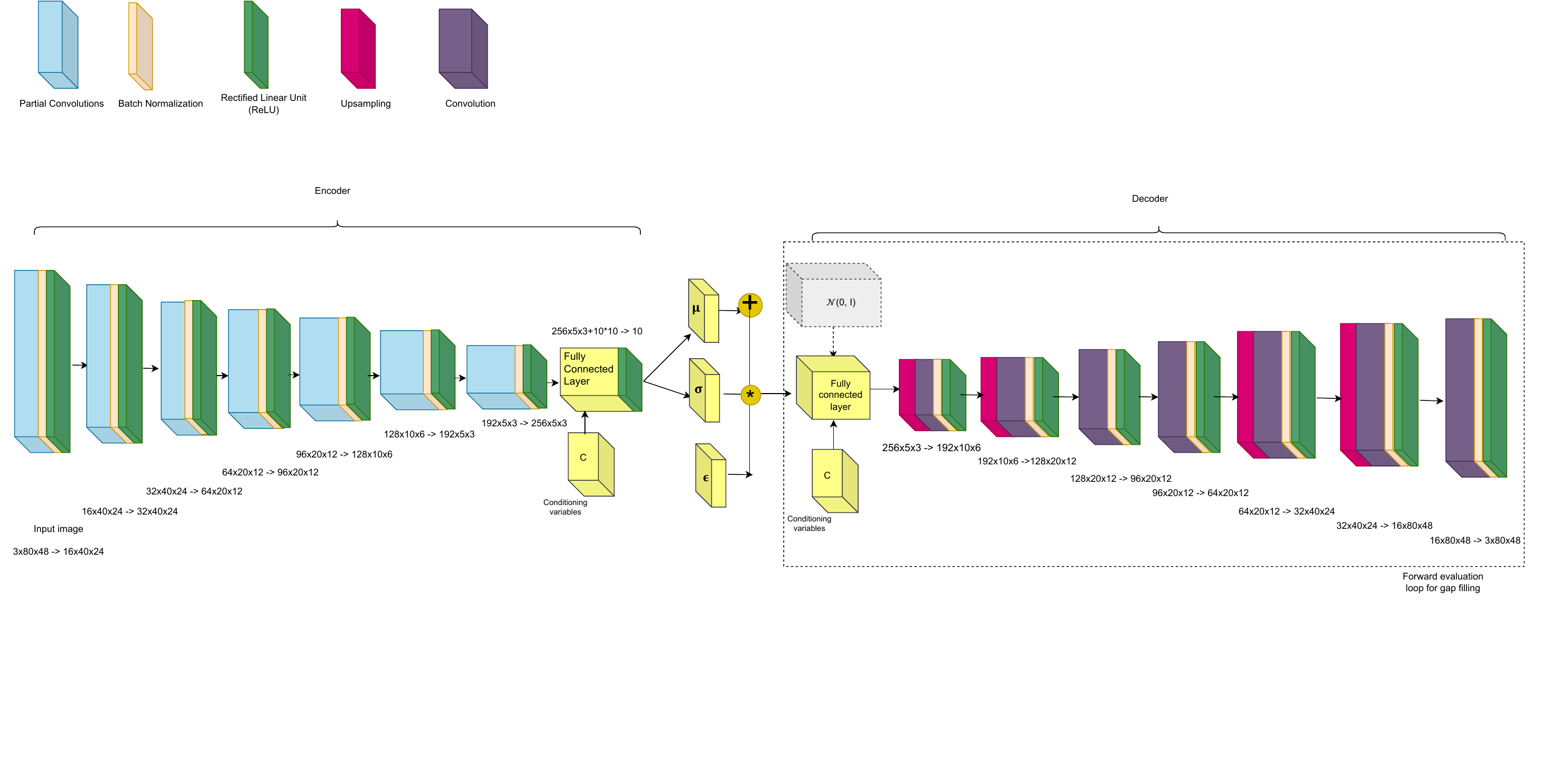}
    \caption{Schematic of the CVAE network used in this study. Forward evaluation loop is highlighted with a box with dashed edges.}
    \label{fig:cvae_network}
\end{figure*}
\subsection{Encoder and Decoder network}
\label{sec:encoderDecoder}
The encoder and the decoder networks in the CVAE are constructed using a series of convolutional neural networks (CNN) and a couple of fully connected layers. CNNs are well-suited for recognizing specific patterns in the input data. In order to accommodate the irregularly shaped gaps in the PV data, a technique known as partial convolutions~\cite{liu2018image} is used to modify the convolution operators. This critical step ensures that only the valid data are used for training the model. The gappy regions are masked off and are not utilized in the training. Figure~\ref{fig:cvae_network} shows the network used in this study. A 3$\times$3 kernel was used to detect spatially fine features. Due to the gaps in the training dataset, partial convolution operators are used in the encoder network. The decoder network contains the regular CNN operators. Batch normalization~\cite{batchnorm2D} and rectified linear unit (ReLU)~\cite{relu} activation function along with fully connected layers are also shown in Fig.~\ref{fig:cvae_network}. For all the results presented in this study, a 10 dimensional latent space was used. This parameter and all the other hyperparameters related to training process such as learning rates were optimized using SHERPA, a python based hyperparameter optimization library~\cite{hertel2020sherpa}.

\subsubsection{Partial convolution}
\label{sec:partialConv}
As mentioned in the previous section, only the data from valid regions are used to train the variational auto-encoder (VAE). The spatial regions where gaps were present are replaced with zeros in each of the snapshot of the PIV dataset. Additionally a binary mask same size as the original data is used to denote regions where vectors are eliminated. In order to not use the eliminated vectors for training, partial convolutional filters~\cite{liu2018image}, instead of regular convolutional filters, are utilized in this study. With partial convolutions, provided a binary mask, only the non-masked valid regions are convolved at each layer of the network. Additionally, a mask update step is performed at each layer providing regions of unmasked values for the next layer.  As described in~\cite{liu2018image}, this partial convolution algorithm, relying only on unmasked values, can be expressed as 

\begin{multline}
  x^{\prime}=\begin{cases}
               \mathbf{W}^{T} \left(\mathbf{X} \odot M\right) \frac{sum(l)}{sum(M)} + b,\\
               ~~~~~~~~~if~~\mathbf{sum}(M) > 0\\
               0, ~~~otherwise
                           \end{cases}
                           \label{eq:partialConv1}
\end{multline}
where $\mathbf{W}$ are the convolution weights, $\mathbf{X}$ are the input data, $M$ are the binary masks, $\odot$ implies element-wise multiplication. The matrix $l$ has the same shape as $M$ but with all elements equal to 1 to provide the right scaling. 
\subsection{Conditioning variables}
From each sample in the temporally resolved PIV dataset, a set of 10 conditioning variables are generated. These conditioning variables only utilize the valid vector field. These conditioning variables are designed to uniquely tag each temporal snapshot of the dataset. 
\begin{align}
    \mathbf{C}^i &= \begin{bmatrix}
           \Delta t * i \\
           ^{min} U_{x}^i \\
           ^{max} U_{x}^i \\
           ^{min} U_{y}^i \\
           ^{max} U_{y}^i \\
           ^{min} U_{z}^i \\
           ^{max} U_{z}^i \\
           \sum U_{x}^i \left(x, y\right) \\
           \sum U_{y}^i \left(x, y\right) \\
           \sum U_{z}^i \left(x, y\right)
         \end{bmatrix}
  \end{align}
  where $\Delta t $ is the temporal resolution of the dataset, $U_{1}^i$, $U_{2}^i$ and $U_{3}^i$ are x, y and z components of $i^{th}$ snapshot, and $^{min} U_{1}^i$ is the minimum of the x component of the $i^{th}$ sample gappy vector field.

\subsection{Loss Function}
Equation~\ref{eq:lossfunc} describes the loss function used in this study for backpropagation using stochastic gradient descent (SGD). The reconstruction loss, $MSE(\mathbf{x}, \mathbf{x}_d)$, is only calculated over the valid vector field which are tagged using the binary mask, $M$. The Adam optimizer~\cite{kingma2014adam}, a variant of SGD, is used in this study. A learning rate of $7.5e-4$ was used network training. 
\subsection{Training and Forward evaluation}
For each PIV dataset with gaps, a tailored CVAE network is trained. Once trained, only a forward evaluation step is required. During training, all time snapshots are randomly shuffled and grouped into batches of 32. The corresponding conditioning variables are also computed and shuffled in the same order. The encoder, utilizing multi-layered CNNs and incorporating conditioning variables, aims to map the increasingly complex spatial patterns in each snapshot, independent of the temporal order, to a unique latent space region. In all the datasets presented in the paper a total of 1000 epochs were used for training the respective models. The validation loss was found to reduce by 2 orders of magnitude within 1000 epochs. 

For a forward evaluation, the latent space (unit normal gaussian) distribution is first sampled and is passed to the decoder network along with the conditioning variables. For each snapshot, 32 independent evaluations are made and a mean was computed over these 32 samples to generate a vector field at that instant. The decoder network will generate a full vector field. However, only the vector field in the gaps are retained. The remaining regions, away from the gaps, are patched with velocity field from the training dataset.
\subsection{Implementation Details}
The python based software code for the algorithm described in this study is implemented using the Pytorch library~\cite{paszke2017automatic}. The code is designed to use GPUs, whenever available, for efficient computation. The forward evaluation to generate 5120 snapshots using the decoder network took 0.1 hours on a NVIDIA Tesla V100. 
\section{Experimental Data}
In order to evaluate the CVAE based algorithm to fill gaps, a PIV dataset with high vector yield and low uncertainty is chosen. Similar to the procedure described in~\cite{saini2016development}, gaps with varying pixel widths are created in the data and vectors in these gaps are completely eliminated. The performance of the CVAE algorithm to fill these gaps is then investigated by comparing against original data without any gaps. 
\label{sec:exp}
\subsection{Premixed Swirl Burner}
\begin{figure}[!htb]
    \centering
    \includegraphics[width=0.4\linewidth]{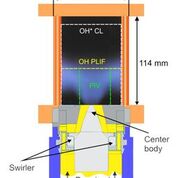}
    \caption{Schematic of the experimental configuration~\cite{an2019role}}
    \label{fig:expConfig}
\end{figure}

In this study, Stereo-PIV data from the experimental study of the premixed swirl burner~\cite{an2019role,meier2007detailed,caux2014thermo} is considered. The experimental setup, shown in Fig.~\ref{fig:expConfig}, was equipped with simultaneous 10 kHz OH planar laser induced fluorescence (PLIF) and stereoscopic particle image velocimetry (S-PIV) diagnostic. The experimental diagnostics and data processing techniques are discussed in detailed in~\cite{an2019role} and will not be discussed in this paper.  
Three different cases corresponding to three different operating conditions were considered in this study. The following table lists out the details of each of these cases. A total of 5120 snapshots for each of the test case was used to train the CVAE network for each of the cases. 

\begin{figure}[!htb]
    \centering
    \includegraphics[width=\columnwidth]{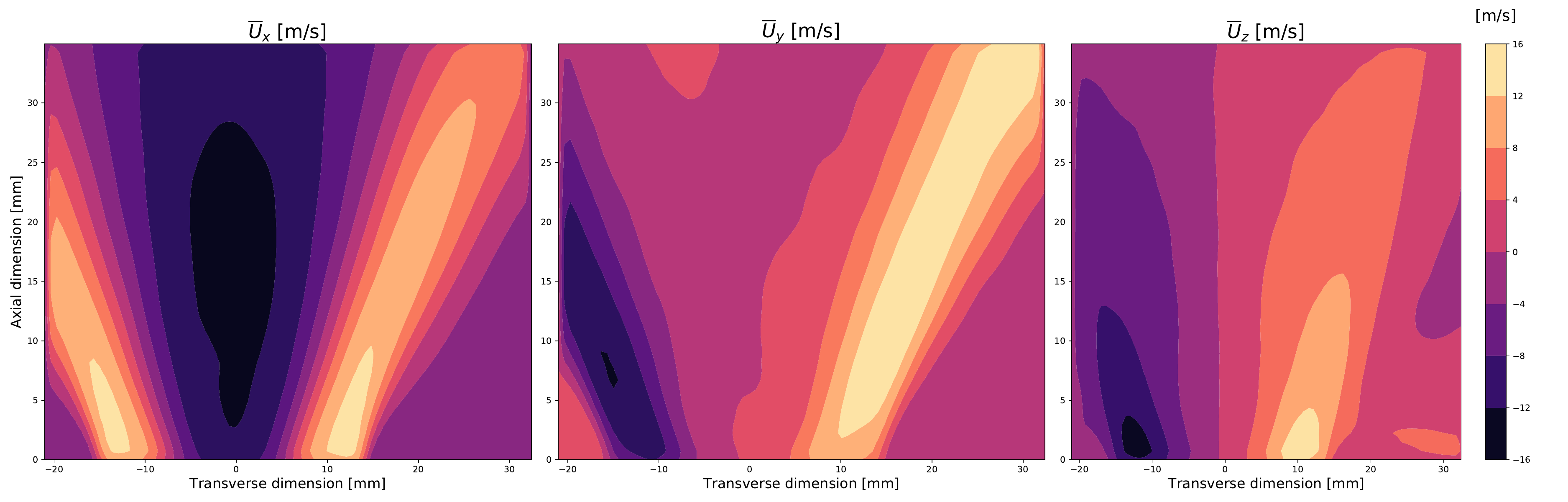}
    \caption{Time averaged velocity field from case D2F3.}
    \label{fig:TimeAvD2F3}
\end{figure}

\begin{table}[h!] \footnotesize
     \caption{Combustor operating conditions for all the cases considered in this study}
 \resizebox{\columnwidth}{!}{\begin{tabular} {|l|l|l|l|l|l|l|l|}
    \hline
         Case name& $T_{ad}$ & $T_r$ & $D_r$ & $\phi$ & $X_{air}$ & $X_{CH_4}$ & $X_{H_2}$  \\ 
         \hline
         D2F3 & 1750 K & 400K & 4.4 & 0.6 & 0.93 & 0.055 & 0.014 \\
         \hline
         D2F1 & 1750K & 400K & 4.4 & 0.6 & 0.94 & 0.060 & 0.0 \\
         \hline
         D3F3 & 1700K & 500K & 3.4 & 0.53 & 0.94 & 0.049 & 0.012\\
         \hline
    \end{tabular}}
    \label{tab:expCondition}
\end{table}
The D in the case name refers to the reactant-to-product density ratio. This parameter is one of the critical parameters impacting hydrodynamic instabilities leading to distinct velocity fields. In this study, two cases with different density ratios, D2 and D3, were considered. The letter F in the case name is to denote the H2 addition in the fuel stream. The H2 fuel addition was shown to play a critical role in increasing flame stability in the experimental study using this rig~\cite{an2019role}. For all the three cases considered in this study, a complex flow field was observed, Fig.~\ref{fig:TimeAvD2F3}. The flow field is characterized by the swirling reactant jet, inner and outer shear layers along with a central recirculation zone (CRZ). For a detailed discussion regarding the flow features and flame stabilization in this burner, the reader is referred to~\cite{an2019role}. The time resolved SPIV (3 velocity components on a 2D plane) data in these experimental measurements had a very high vector yield with low uncertainty. As is typically the case with S-PIV data the out of plane velocity component, $U_z$, was found to have the highest uncertainty. Owing to the high level of confidence in the vector field obtained from this experimental study, this dataset was selected for studying the efficacy of CVAE in generating velocity fields within gaps and comparing them to the original data in those artificial gaps. The algorithm for creating these artificial gaps will be discussed further.
\subsection{Algorithm for vector elimination}
\begin{figure}[!htb]
    \centering
    \includegraphics[width=0.5\columnwidth]{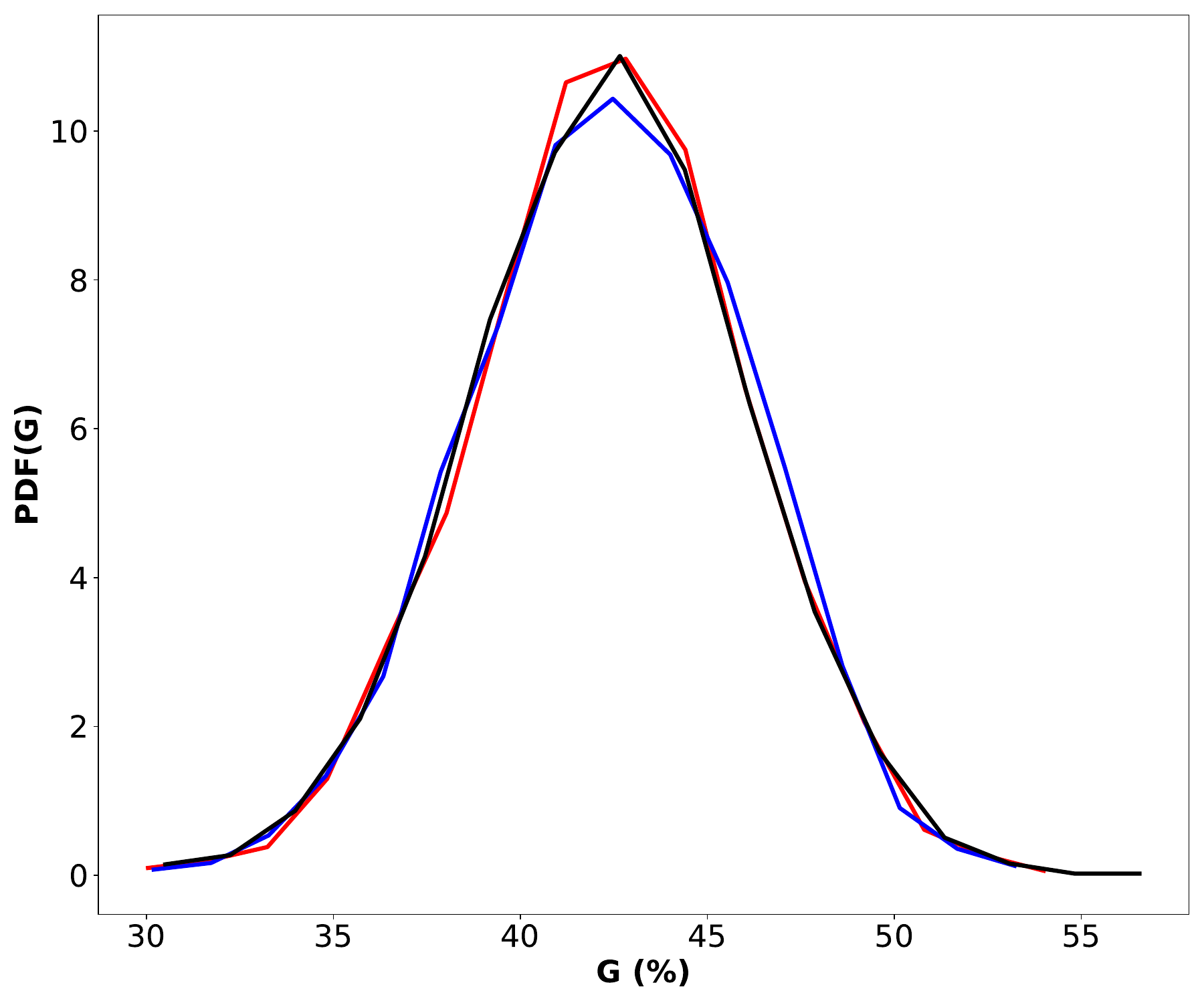}
    \caption{Probability density function (PDF) of \% of vectors eliminated in each of the 5120 images. Red line corresponds to the case D2F1, blue line is from case D2F3 and black line corresponds to case D3F3.}
    \label{fig:coverage}
\end{figure}

Only 50\% and 75\% gappiness percentage (G), ratio of missing vectors to the total number of possible vectors, are considered in this study. For the G$\approx$50\% case the total surface area of the gaps in the time-resolved data are distributed across all snapshots following a Gaussian profile in time, with the mean value of gappiness denoted by G. In each snapshot, these artificial gaps are randomly distributed throughout the spatial domain. The algorithm used in the gappy POD study~\cite{saini2016development} had a restriction of no single point in the field was missing for all snapshots. This constraint has been removed in the gap creation algorithm used in this study. Figure~\ref{fig:coverage} shows the distribution of the \% of vector eliminated from each of the snapshot spread across the entire 5120 snapshots in the dataset from each of the three cases.
\section{Results and Discussion}
\label{sec:results}
To evaluate the effectiveness of the proposed CVAE algorithm, three distinct networks were trained for each of the operating conditions, Tab.~\ref{tab:expCondition}. For each of the operating condition, a distinct CVAE network is trained using a dataset with a mean G$\approx$50\%, Fig.~\ref{fig:coverage}. The efficacy of CVAE will also be assessed on a test case with G$\approx$75\% and will be presented at the end of this section. Such a significant data loss in a realistic PIV dataset will warrant a significant change in the experimental methodology along with instrumentation design and therefore will rarely be seen. However, this study is focused on stretching the limits of the proposed CVAE technique and providing a usable guide to the user to help them make an informed judgement on CVAE for gap filling. For this reason a case with G$\approx$75\% is also being considered in this study. In terms of metrics to evaluate the performance of the proposed technique, first a qualitative comparison between the original 2D vector field and the reconstructed field will be shown. Thereafter, reconstruction error (E), defined as below and also used in~\cite{saini2016development} will be presented. 
\begin{equation}
    E = \sqrt{\sum(U^{r} - U^{o})^2 / \sum (U^o)^2}
    \label{eq:reconError}
\end{equation}
Statistics of spatially local relative error, $\mid\frac{U^{r}-U^{o}}{U^{o}}\mid$ where $U^{r}$ is the reconstructed field and the original field is represented by $U^{o}$, will also be presented.  

\begin{figure*}[!htb]
    \centering
    \includegraphics[width=0.9\textwidth]{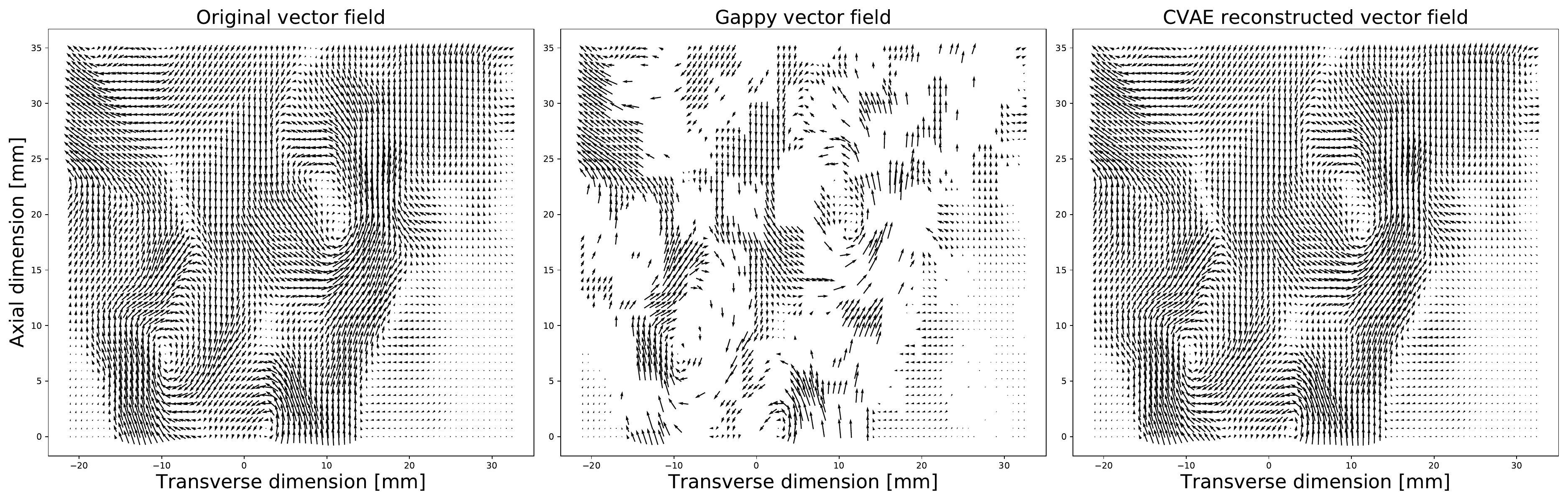}
    \caption{Instantaneous vector plot showing the original vector field, vector field with 55\% of vectors eliminated, CVAE reconstructed vector field from the case D2F1.}
    \label{fig:vectorPlotD2F1}
\end{figure*}

\begin{figure*}[!htb]
    \centering
    \includegraphics[width=0.9\textwidth]{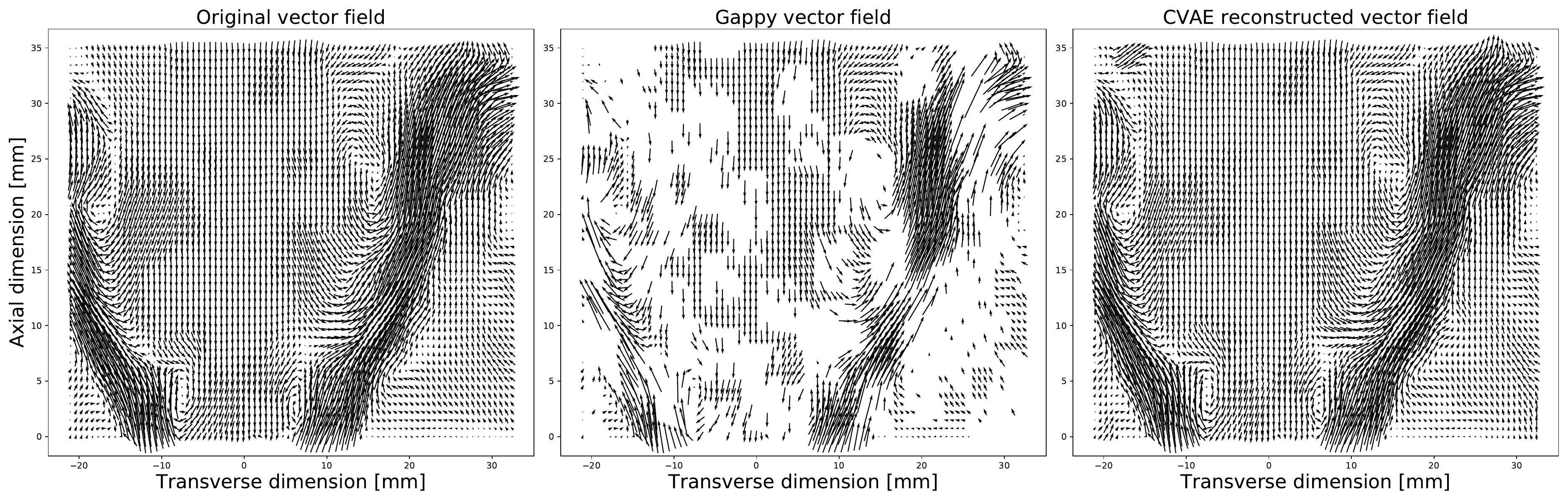}
    \caption{Instantaneous vector plot showing the original vector field, vector field with 54\% of vectors eliminated, CVAE reconstructed vector field from the case D2F3.}
    \label{fig:vectorPlotD2F3}
\end{figure*}

\begin{figure*}[!htb]
    \centering
    \includegraphics[width=0.9\textwidth]{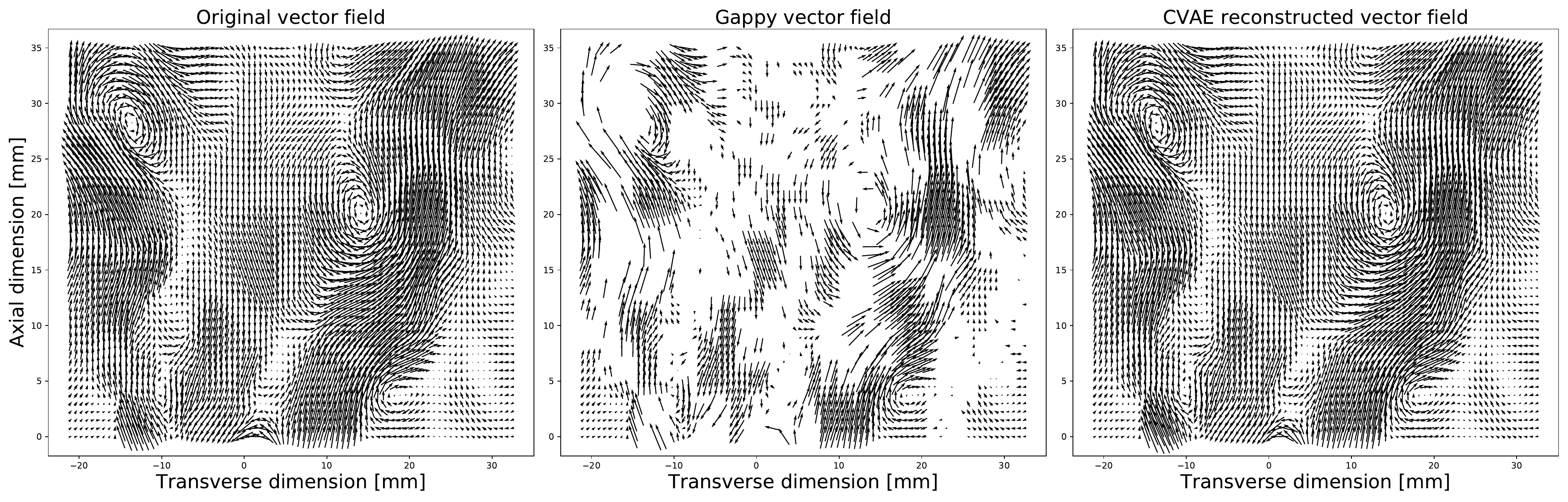}
    \caption{Instantaneous vector plot showing the original vector field, vector field with 58\% of vectors eliminated, CVAE reconstructed vector field from the case D3F3.}
    \label{fig:vectorPlotD3F3}
\end{figure*}

Figures~\ref{fig:vectorPlotD2F1},~\ref{fig:vectorPlotD2F3} and~\ref{fig:vectorPlotD3F3} shows the original clean vector field, vector field with gaps and the reconstructed vector field using the CVAE technique for all the three operating conditions, respectively. The snapshots shown in Fig.~\ref{fig:vectorPlotD2F1},~\ref{fig:vectorPlotD2F3} and~\ref{fig:vectorPlotD3F3} correspond to a time instant where 55\%, 54\% and 58\%, respectively, of vectors were fully eliminated. 

Qualitatively, all the main flow features eliminated during the gap creation was reconstructed accurately using the CVAE strategy. Despite the elimination of large chunks of vector field along swirling arms of the reactant jet in Fig.~\ref{fig:vectorPlotD2F3}, CVAE is able to reproduce these flow features effectively. In Figs.~\ref{fig:vectorPlotD2F1} and~\ref{fig:vectorPlotD3F3} some of the low speed vortical structures, eliminated in the training data, are successfully reconstructed by the CVAE. 

\begin{figure}[ht]%
 \centering
 \subfloat[]{\includegraphics[width=0.5\columnwidth]{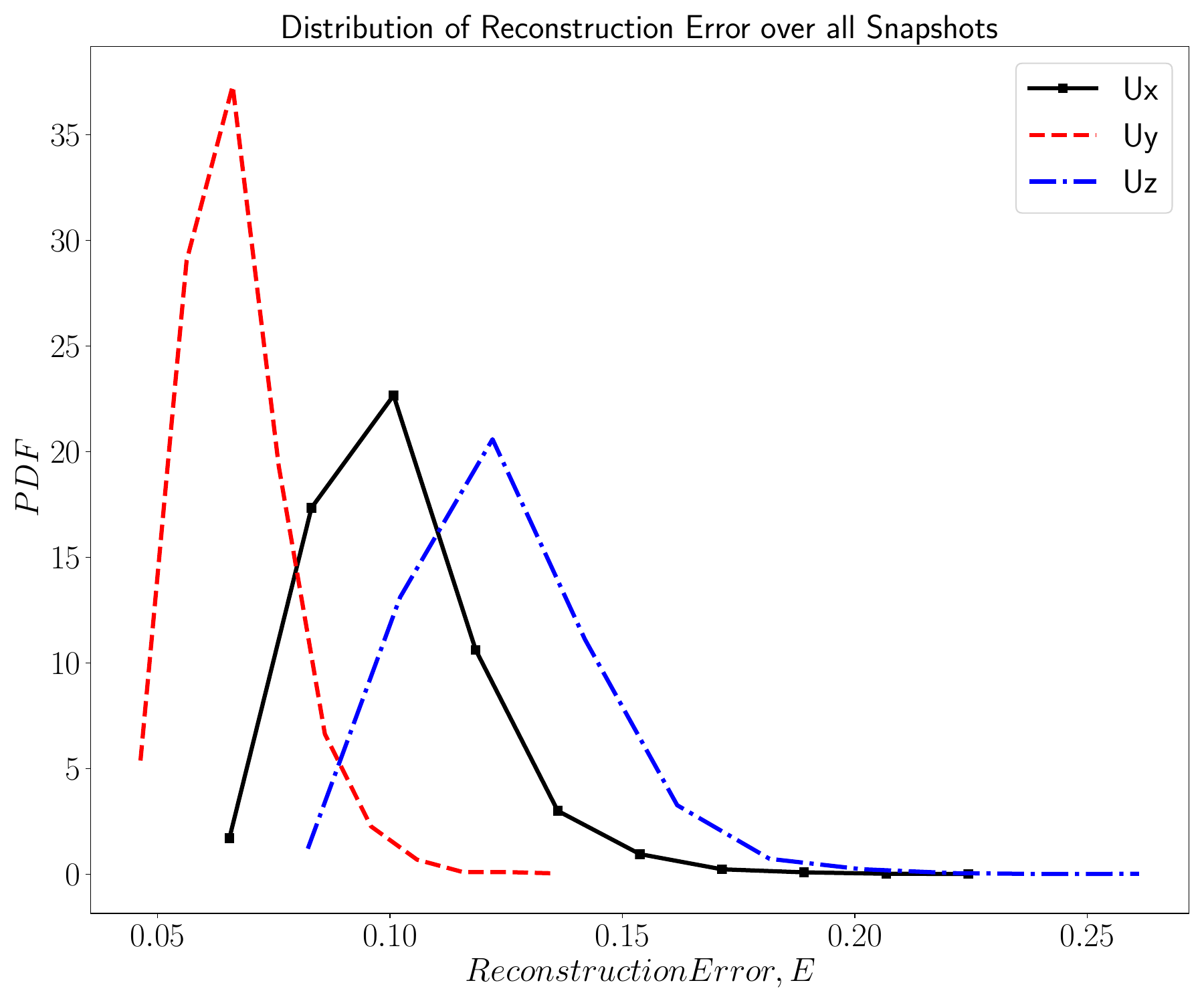}}\label{fig:D2F1_FrobError}%
 \subfloat[]{\includegraphics[width=0.5\columnwidth]{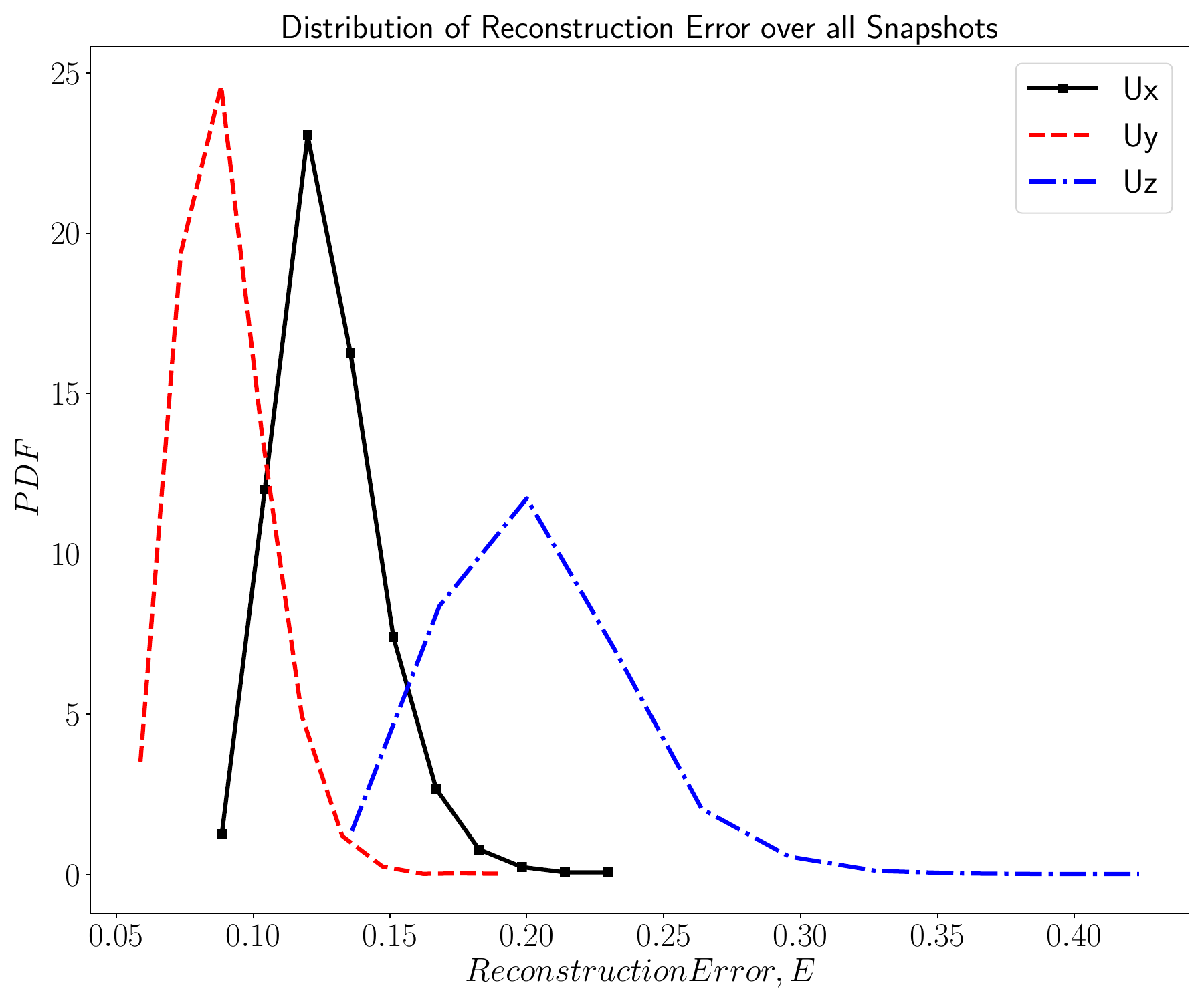}}\label{fig:D2F3_FrobError}\\
 \subfloat[]{\includegraphics[width=0.5\columnwidth]{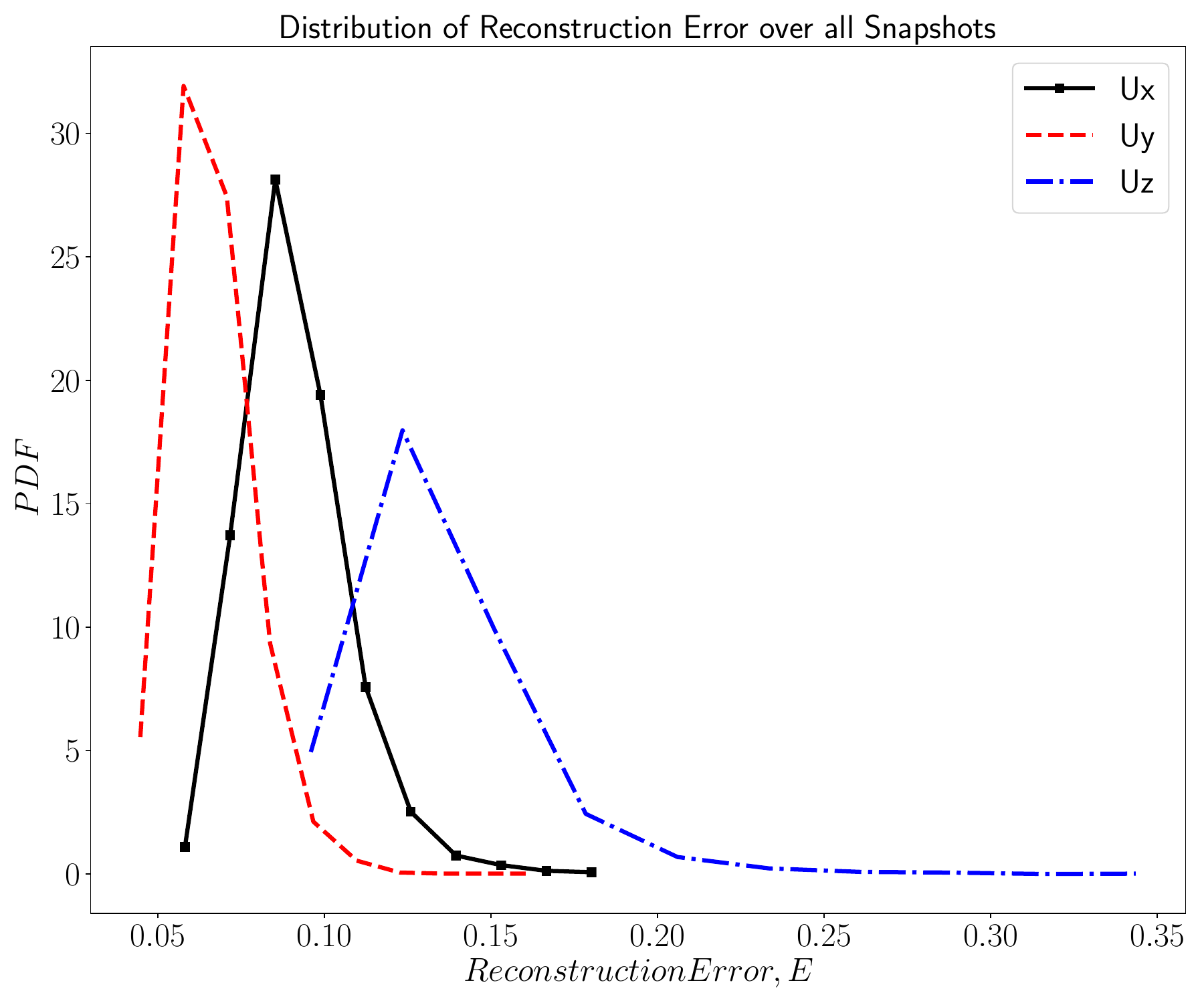}}\label{fig:D3F3_FrobError}%
 \caption{PDF, computed over all the snapshots, of the reconstruction error, E from the three test cases considered in this study.(a) Case D2F1, (b) Case D2F3, (c) Case D3F3}%
 \label{fig:froError}%
\end{figure}

\begin{figure}[ht]%
 \centering
 \subfloat[]{\includegraphics[width=0.5\columnwidth]{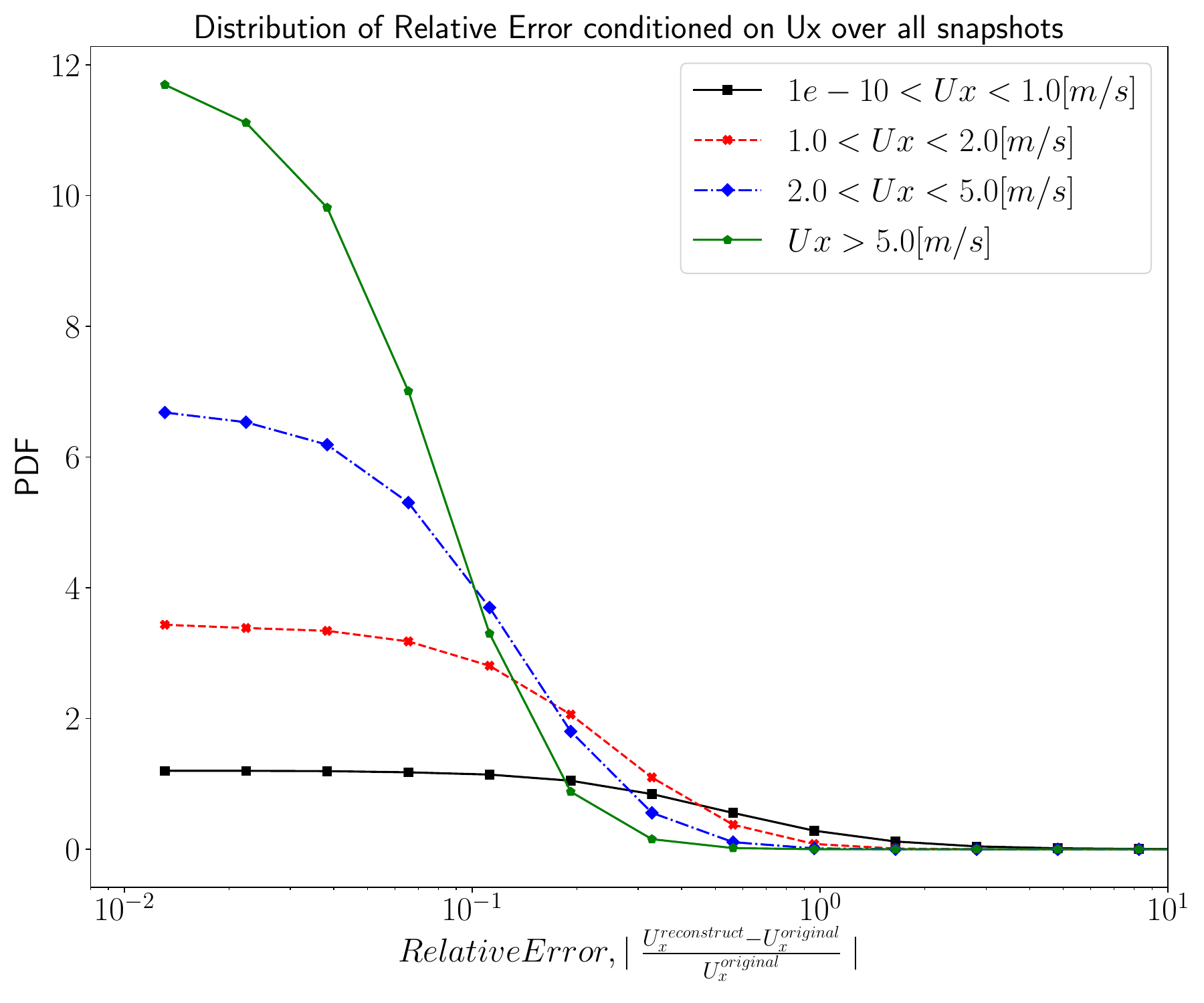}}\label{fig:D2F1_RelError}%
 \subfloat[]{\includegraphics[width=0.5\columnwidth]{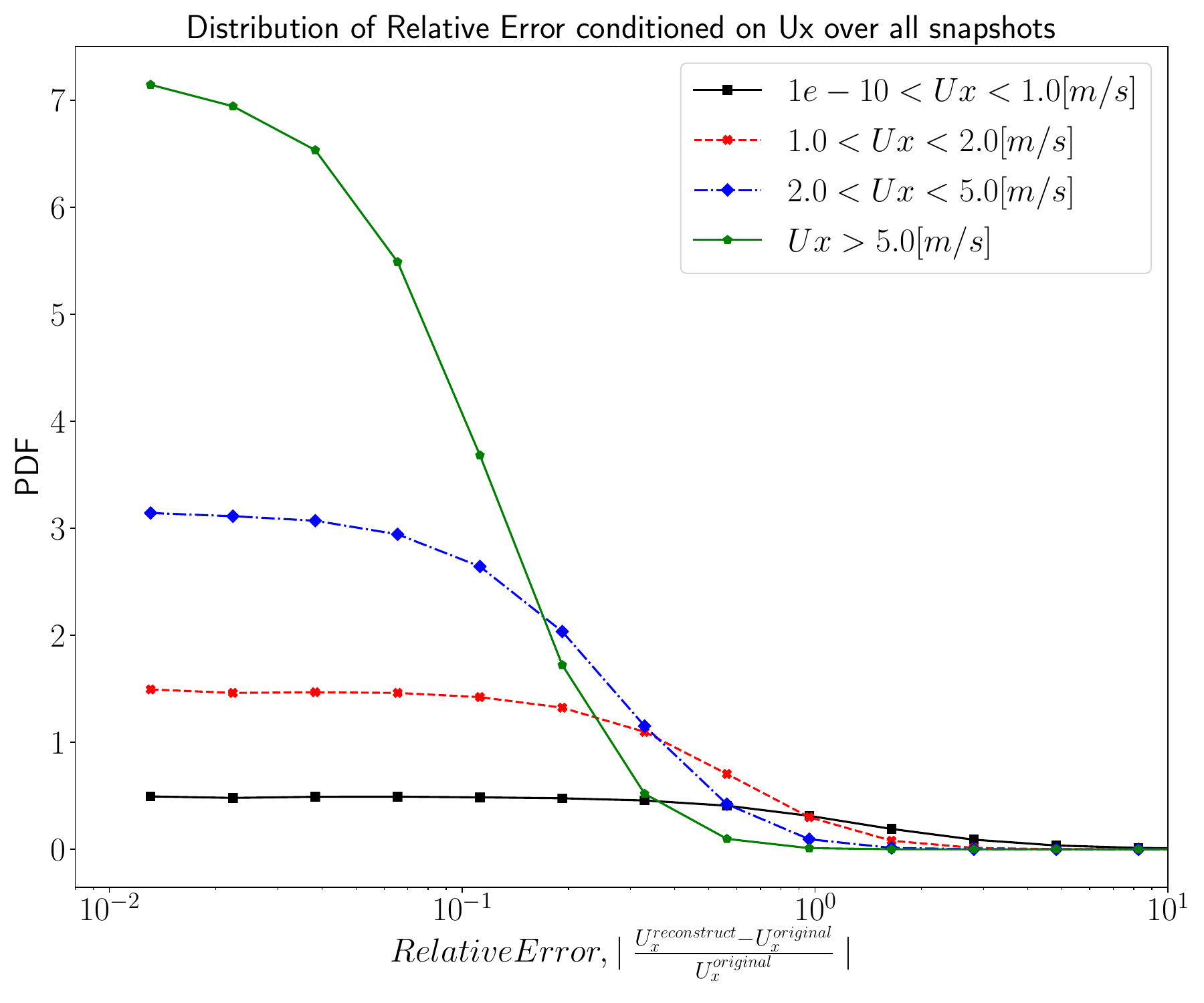}}\label{fig:D2F3_RelError}\\
 \subfloat[]{\includegraphics[width=0.5\columnwidth]{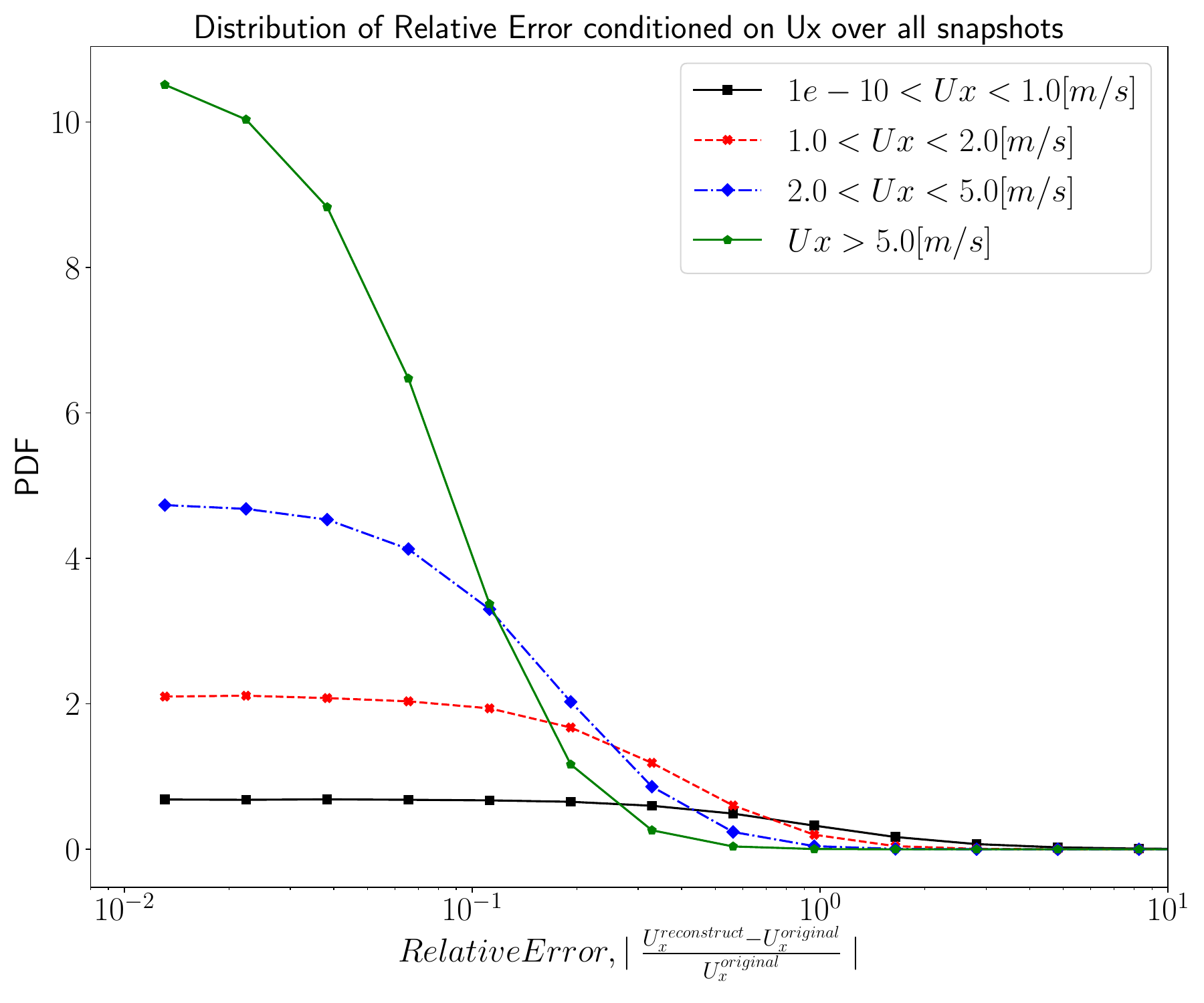}}\label{fig:D3F3_RelError}%
 \caption{PDF, computed over an entire snapshot and over all the snapshots in the dataset, of the relative error $\mid\frac{U^{r}_x-U^{o}_x}{U^{o}_x}\mid$ of $U_x$ conditioned on a range of $U_x$ values.(a) Case D2F1, (b) Case D2F3, (c) Case D3F3}%
 \label{fig:relError}%
\end{figure}

Figure~\ref{fig:froError} shows the distribution of reconstruction error, E, for all three components of velocity over all the snapshots for each of the test case. The reconstruction error, Eq.~\ref{eq:reconError} without the term in the denominator, is used in the loss function (Eq.~\ref{eq:lossfunc}) used to train the network. Figure~\ref{fig:froError} shows that $U_y$ velocity component has the lowest error amongst all the components followed by $U_x$ and $U_z$.  This trend is similar to the trend presented in~\cite{saini2016development}. The higher uncertainty and the noise in the out-plane velocity components in SPIV~\cite{prasad2000stereoscopic} was attributed by the GPOD study~\cite{saini2016development} as main cause of higher error in the $U_z$ component. By comparing the reconstruction errors obtained by CVAE against the various GPOD variants presented in~\cite{saini2016development}, it is evident that CVAE achieves lower reconstruction errors. Moreover, this superior performance of CVAE is attained using a training dataset with a reduced number of valid vectors.
To analyze the sources of these errors distribution of spatially local relative error defined as, $\mid\frac{U^{r}_x-U^{o}_x}{U^{o}_x}\mid$, is shown in Fig.~\ref{fig:relError}. This distribution of relative error, in the $U_x$ component, is computed over all points and over all the snapshots. To highlight the regions where this error is observed to be the highest, the relative error is conditioned on the local $U_x$. Figure~\ref{fig:relError} shows that the regions with higher velocity have smaller relative errors whereas the regions with negligible velocities dominate the relative error metric. For the highest velocity zones CVAE predicted a relative error $<0.2$ in the $U_x$ velocity component.  The same trend in relative error is observed for all the cases.

\begin{figure}[!htb]
    \centering
    \includegraphics[width=0.75\columnwidth]{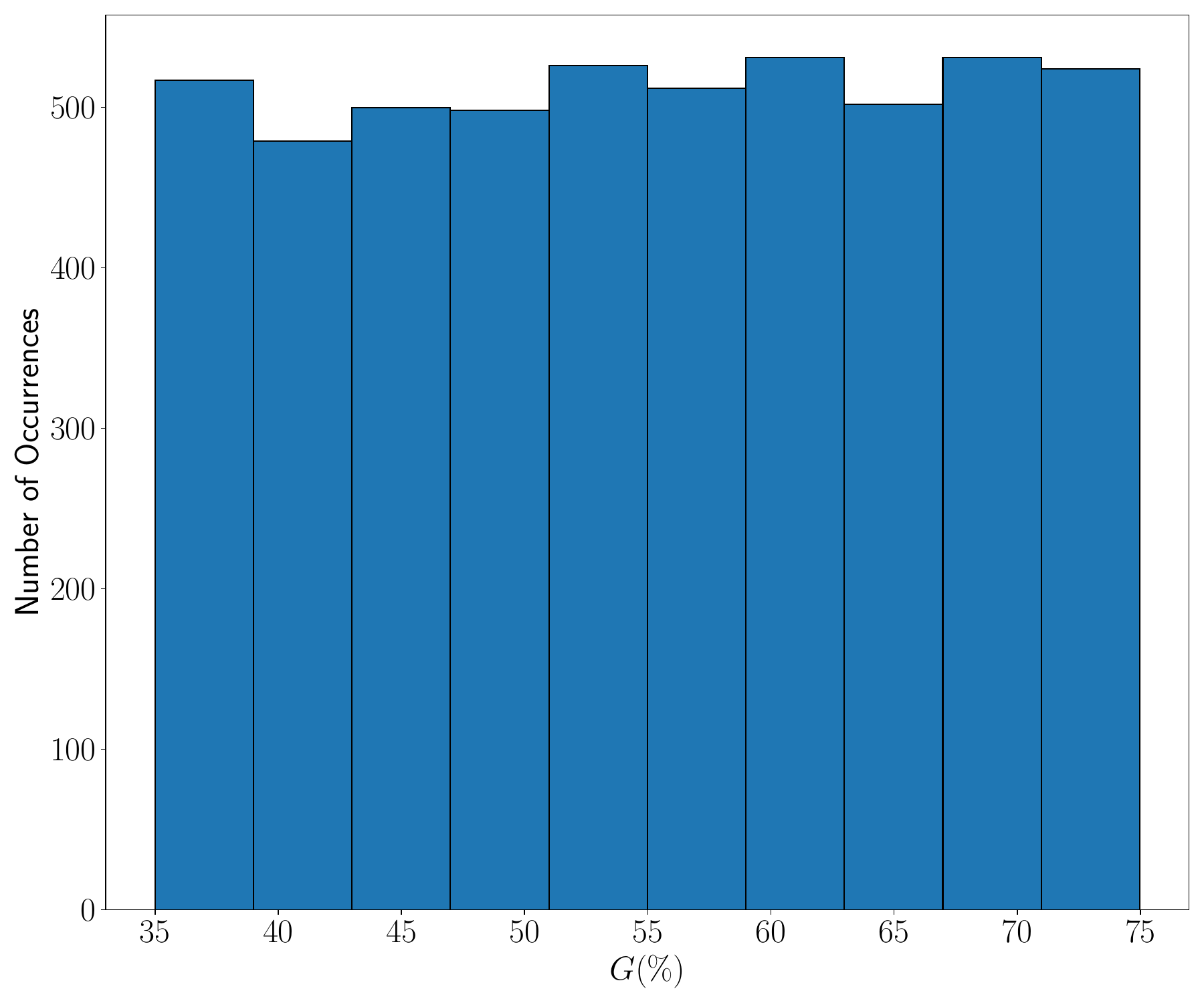}
    \caption{Histogram showing number of snapshots with \% of vectors eliminated from the dataset corresponding to the case D2F1}
    \label{fig:coverage75}
\end{figure}

\begin{figure*}[!htb]
    \centering
    \includegraphics[width=\textwidth]{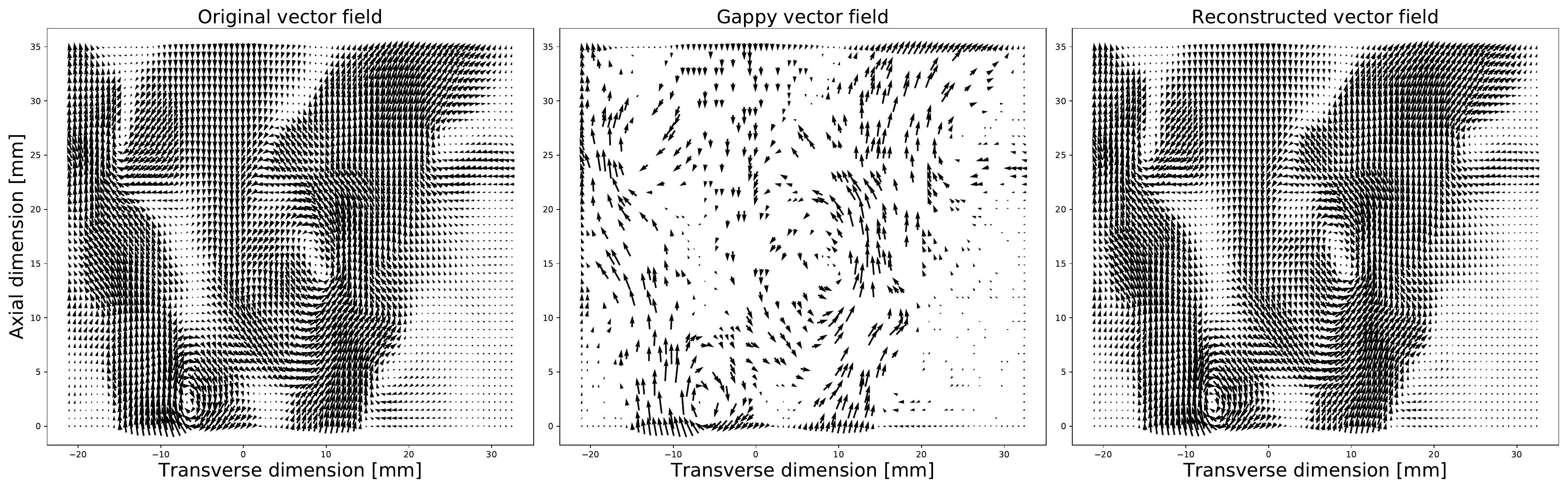}
    \caption{Instantaneous vector plot showing the original vector field, vector field with 75\% of vectors eliminated, CVAE reconstructed vector field from the case D2F1.}
    \label{fig:vectorPlotD2F175}
\end{figure*}

\begin{figure}[!htb]
    \centering
    \includegraphics[width=0.75\columnwidth]{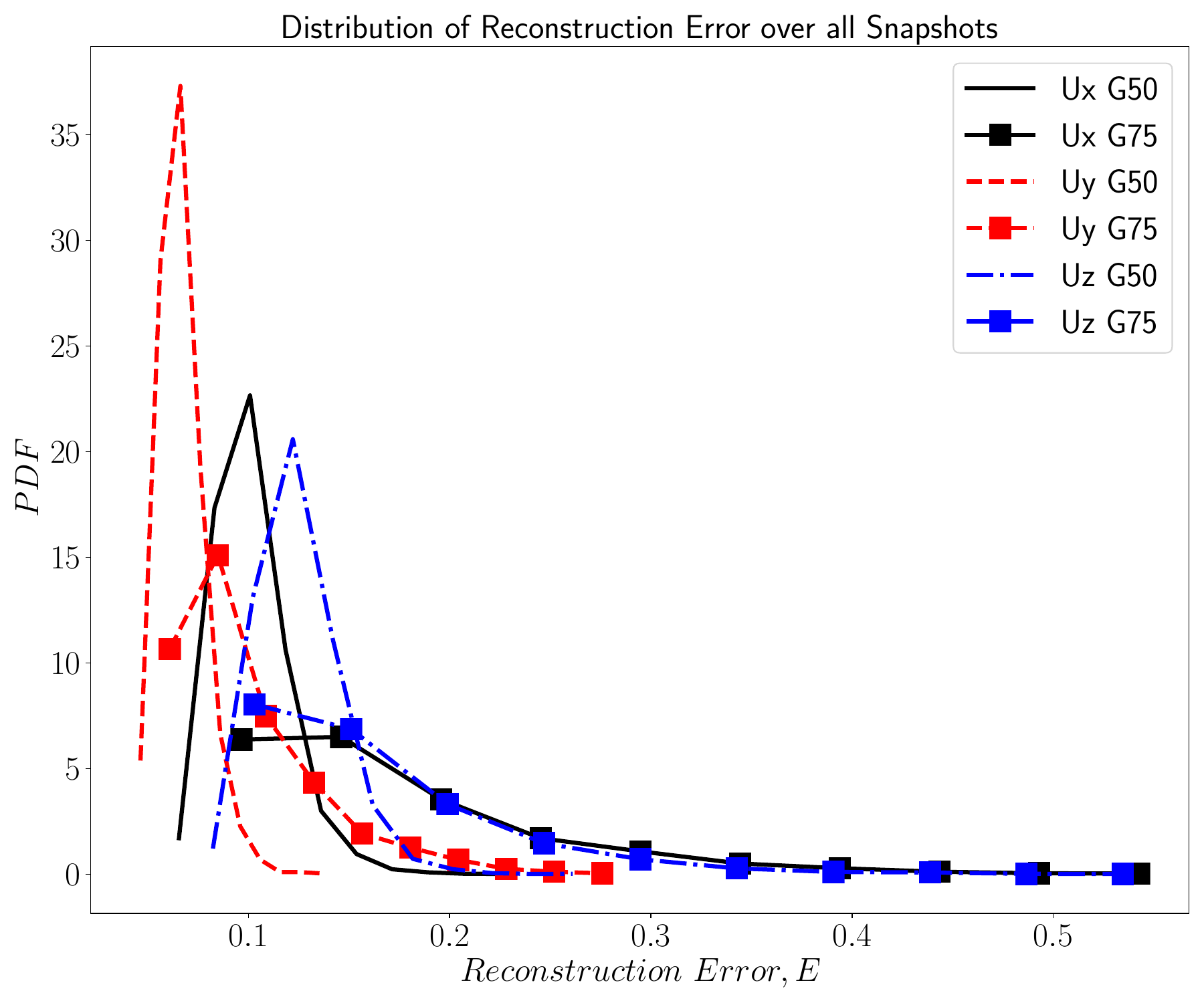}
    \caption{PDF, computed over all the snapshots, of the reconstruction error, E. Lines are from the case with G 50\% and lines with symbols are from the case with G 75\%.}
    \label{fig:frobError5075}
\end{figure}

To demonstrate the capability of CVAE in accurately reproducing velocity field in gaps even in extreme cases, a test case is designed with some snapshots with 75\% of vectors eliminated. For this case the training dataset was generated with a uniform distribution of \% of vectors eliminated, ranging between 35-75\%, spread uniformly over all the 5120 snapshots, Fig.~\ref{fig:coverage75}.  The dataset for the case D2F1 was used to generate the gappy training set. A distinct CVAE network was trained with this dataset with gaps. Figure~\ref{fig:vectorPlotD2F175} shows the instantaneous original vector field, field with artificial gaps and the reconstructed field using CVAE for a snapshot where 75\% of the vectors were eliminated from the training set. Remarkably, the CVAE latent space along with the decoder network were able to reproduce all the flow features seen in the original vector field. Figure~\ref{fig:frobError5075} shows the reconstruction error using the CVAE trained with G$\approx$ 75\% dataset and compares against the prediction made by the network trained using the D2F1 dataset with G$\approx$ 50\%. As expected higher reconstruction errors are predicted with the CVAE network trained on snapshots with gappy \% shown in Fig.~\ref{fig:coverage75}. However, even in this extreme case reconstruction errors predicted by CVAE are similar to what was observed with the MF GPOD method described in~\cite{saini2016development}.

\section{Conclusions}
\label{sec:conclusion}
A deep learning-based generative technique called the Conditional Variational Auto-Encoder (CVAE) is proposed for reconstructing velocity vector field in gaps typically observed in reacting particle image velocimetry (PIV) data. This non-linear dimensionality reduction technique, characterized by its superior latent space properties, solely relies on the valid vector field and is demonstrated to accurately reproduce vector field in the gaps. The qualitative and quantitative performance of the CVAE technique is demonstrated using stereo-PIV data from a premixed swirl combustor with artificial gaps where up to 75\% of the vector field was removed from some of the snapshots. This highly versatile technique also offers excellent data compression properties enabling the efficient processing of large-scale PIV data.

\acknowledgement{Declaration of competing interest} \addvspace{10pt}

The author declares that he has no known competing financial interests or personal relationships that could have appeared to influence the work reported in this paper.

\acknowledgement{Acknowledgments} \addvspace{10pt}
The author would like to gratefully acknowledge Adam Steinberg and Pankaj Saini for sharing the SPIV dataset and source codes for generating gappy PIV fields. 

This research was supported by the Exascale Computing Project (ECP), Project Number: 17-SC-20-SC, a collaborative effort of two DOE organizations - the Office of Science and the National Nuclear Security Administration, responsible for the planning and preparation of a capable exascale ecosystem, including software, applications, hardware, advanced system engineering and early testbed platforms, to support the nation's exascale computing imperative. The research was performed using computational resources sponsored by the Department of Energy's Office of Energy Efficiency and Renewable Energy and located at the National Renewable Energy Laboratory.





 \footnotesize
 \baselineskip 9pt


\bibliographystyle{pci}
\bibliography{PCI_LaTeX}


\newpage

\small
\baselineskip 10pt



\end{document}